\def\ga{\gamma}
\def\de{\delta}
\def\ep{\epsilon}

\def\la{\lambda}

\def\si{\sigma}

\def\ph{\phi}

\def\ps{\psi}

\def\Ph{\Phi}

\def\cF{{\cal F}}
\def\cL{{\cal L}}

\def\frac#1#2{\textstyle{{{#1} \over {#2}}}}

\def\prt{\partial}

\def\half{{\textstyle{1\over 2}}}
\def\lsim{\mathrel{\rlap{\lower4pt\hbox{\hskip1pt$\sim$}}
    \raise1pt\hbox{$<$}}}
\def\gsim{\mathrel{\rlap{\lower4pt\hbox{\hskip1pt$\sim$}}
    \raise1pt\hbox{$>$}}}
\def\ol#1{\overline{#1}}

\def\slasha#1{\setbox0=\hbox{$#1$}#1\hskip-\wd0\hbox to\wd0{\hss\sl/\/\hss}}
\def\slashb#1{\setbox0=\hbox{$#1$}#1\hskip-\wd0\dimen0=5pt\advance
       \dimen0 by-\ht0\advance\dimen0 by\dp0\lower0.5\dimen0\hbox
         to\wd0{\hss\sl/\/\hss}}

\documentclass[
    ,final            
  ]
  {aipproc}

\layoutstyle{6x9}

\begin{document}

\hbox to \hsize{
\hfill\vbox{\hbox{\bf IUHET-454}
            \hbox{December 2002}} }

\title{Lorentz Violation and Spacetime Supersymmetry\footnote{Presented at 
Coral Gables Conference: Short Distance Behavior of Fundamental Interactions,
December 11-14, 2002, Fort Lauderdale, Florida.}}

\author{M. S. Berger}{
  address={Physics Department, Indiana University, Bloomington, 
IN 47405, USA\\E-mail: berger@indiana.edu}
}

\begin{abstract}
Supersymmetry and Lorentz invariance are closely related
as both are spacetime symmetries. Terms can be added to Lagrangians that 
explicitly break either supersymmetry or Lorentz invariance. It is possible to 
include terms which violate Lorentz invariance but maintain invariance 
under supersymmetric transformations. I illustrate this 
with some simple extensions of the original Wess-Zumino model.
\end{abstract}

\maketitle


\section{Introduction}

The understanding of spacetime symmetries has grown remarkably in the last 
century. The mass-momentum behavior is associated with the requirement 
that theories respect translation invariance, while the intrinsic spin and 
the existence of antiparticles are consequences of the Lorentz group.
The theoretical discovery of supersymmetry is a major
achievement of twentieth century physics, and it marked 
another expansion of our concept of spacetime symmetries.
The advent of supersymmetry was followed by its application to gauge
theories and the Standard Model and a local version (supergravity) that 
incorporates general relativity. This is a clear indication that supersymmetry
and Lorentz invariance are closely linked.
The introduction of fermionic generators that relate fermions and bosons 
resulted largely from motivations that were theoretical, but it is clear that
supersymmetry has come to dominate particle physics phenomenology. 

A large part of the history of physics and particle physics in particular 
involves the introduction of symmetries and the understanding of the 
spontaneous breaking of these symmetries. The longstanding 
point of view has been that 
spacetime symmetry is one of the most fundamental.
The familiar symmetries regarding translations,
angular momentum, and Lorentz boosts that constitute the Poincar\'e group
are usually assumed to be unbroken 
symmetries. While there is great theoretical appeal for these symmetries
to be unbroken, this issue is of course a question that can only be decided 
by experiment. In any case the amount of breaking of the Lorentz
symmetry must be very small, if it is indeed nonzero,
since it has so far escaped experimental detection.
Supersymmetry, if it exists, comprises another part of the overall
spacetime symmetry, and it is clearly broken. In fact, it must be broken 
badly compared to the scale at which
we perform experiments, so much so that we have currently not detected any 
of the supersymmetric partners to the Standard Model particles. However, 
from a more theoretical point of view, the breaking of supersymmetry is 
also very small since the breaking scale is very much suppressed in 
comparison to the Planck scale (where the true nature of spacetime 
presumably emerges). 

The Poincar\'e algebra involves the generator of translations ($P_\mu$) and
the generator of rotations and Lorentz boosts $M_{\mu \nu}$ in the following 
way
\begin{eqnarray}
\left [ P_\mu,P_\nu \right ] &=&0\nonumber \\
\left [ P_\mu,M_{\rho \sigma}\right ]&=&i(\eta _{\mu \rho}P_\sigma
-\eta _{\mu \sigma}P_\rho)\nonumber \\
\left [ M_{\mu \nu}, M_{\rho \sigma}\right ] &=& i(\eta _{\nu \rho}
M_{\mu \sigma}-\eta _{\nu \sigma}M_{\mu \rho}
-\eta _{\mu \rho}M_{\nu \sigma}
+\eta _{\mu \sigma}M_{\nu \rho})\;,
\label{poincare}
\end{eqnarray}
Explicit terms can be added to a Lagrangian that violate the Poincar\'e
algebra. If one wants to preserve energy-momentum conservation, then any 
Lorentz violation continues to respect the first of these equations.
The set of terms which can be added to the Standard Model have been 
categorized~\cite{Colladay:1997iz,Colladay:1998fq} in an extended Lagrangian. 
These terms are thought to arise in a more fundamental theory like string 
theory~\cite{Kostelecky:1990nt,Kostelecky:1996qk} which is nonlocal, 
but in the context of the Standard Model extension
they are viewed as phenomenological parameters.

Supersymmetry involves extending this algebra with a fermionic generator 
$Q$\footnote{The four-component notation for spinors
makes the relationship with the Standard Model Lorentz-violating 
extension ~\cite{Colladay:1997iz,Colladay:1998fq} transparent.},
\begin{eqnarray}
\left [Q,P_\mu \right ] &=&0\nonumber \\
\left \{Q,\overline{Q}\right \}&=&2\gamma ^\mu P_\mu \;.
\label{susy}
\end{eqnarray}
This part of the algebra is not respected by supersymmetry breaking terms
in a Lagrangian. The breaking is usually required to be soft meaning that 
quadratic divergences continue to be avoided even though supersymmetry is
explicitly broken. It is well-known that, for the required terms to be soft, 
they must be superrenormalizable, and any nonrenormalizable terms are expected
to be suppressed by powers of the Planck mass
$M_P^{-1}$ or some other cutoff scale 
associated with new physics. 
While motivated by more fundamental theories and 
arising from some mechanism of 
spontaneous breaking of supersymmetry presumably in some
nonperturbative sector that is ``hidden'' by making it completely chargeless 
under the Standard Model gauge group. One can abandon any attempt to 
derive these supersymmetry breaking terms from a more fundamental theory, add
them to a supersymmetric model in the most general way, and then view them 
in a purely phenomenological way. 

For the purposes of this talk, when I say a theory is supersymmetric,
I mean that the relevant Lagrangian respects some modified version of the 
transformations in 
Eq.~(\ref{susy}) but that some portion of Poincar\'e transformations
in Eq.~(\ref{poincare}) is not respected.
Specifically, the models presented here involve violations associated
with the Lorentz generator $M_{\mu \nu}$, so that one can say the models 
respect both supersymmetry and translation invariance.
Previous discussion of the superPoincar\'e algebra has been 
with regard to 
its exact realization, or to cases where the subalgebra involving the 
fermionic generator $Q$ is broken. 
 
This talk is based on work done with V.~Alan~Kosteleck\'y~\cite{Berger:2001rm}.

\section{Lorentz-violating Wess-Zumino Model}

The first four-dimensional supersymmetric field theory was written down 
by Wess and Zumino~\cite{Wess:1974tw}. A modest goal is to determine whether
it is possible to have an unbroken supersymmetry even in the presence 
of broken Lorentz symmetry. Can terms that explicitly violate the
Lorentz symmetry be added to the Wess-Zumino model 
that still preserve some version of the 
supersymmetry? This approach of adding explicit terms is in the same 
spirit as the Lorentz-violating Standard Model extension in which 
Lorentz violation is added to the Standard Model\footnote{Local field theories 
that are derived from underlying nonlocal theories 
have well-known problems with causality and positivity. These problems can 
emerge at a high-energy scale determined by the Planck 
mass~\cite{Kostelecky:2001rh}. The correct point of view is to regard these 
field theories as effective theories, and the inconsistencies will 
be reconciled when the full underlying fundamental theory is taken into 
account. In fact such arguments can be 
used to establish upper bounds on the size of the Lorentz violation.}.
Consider the following
Lagrangian,
\begin{eqnarray}
\cL &=&{1\over 2}\partial _\mu A\partial ^\mu A 
+{1\over 2}\partial _\mu B\partial ^\mu B 
+{1\over 2}i\overline{\psi}\slasha{\partial}\psi+{1\over 2}F^2+{1\over 2}G^2
\nonumber \\
&&+m\left (-{1\over 2}\overline{\psi}\psi+AF+BG\right )\nonumber \\
&&+{g\over {\sqrt 2}}
\left (F(A^2-B^2)+2GAB-\overline{\psi}(A-i\gamma _5B)\psi\right )
\nonumber \\
&&+k_{\mu \nu}\partial ^\mu A\partial ^\nu A 
+k_{\mu \nu}\partial ^\mu B\partial ^\nu B 
+{1\over 2}ik_{\mu \nu}\overline{\psi}\gamma^\mu \partial ^\nu \psi\nonumber \\
&&+{1\over 2}k_{\mu \nu}k^\mu _{\:\:\: \rho}(\partial ^\nu A\partial ^\rho A
+\partial ^\nu B\partial ^\rho B)\;,
\label{Lagrangian}
\end{eqnarray}
where one recognizes the first three lines as the original
Wess-Zumino model. The last two lines involve the Lorentz-violating
coefficients $k_{\mu \nu}$ either linearly or quadratically. 
These couplings are simply numbers (with Lorentz 
index labels) that do not change under a particle Lorentz transformation,
which boosts or rotates local field configurations within a fixed 
inertial frame.
Since there are an even number of indices these terms
do not violate the CPT invariance. Without loss of generality, $k_{\mu \nu}$
can be taken to be a real symmetric, traceless, coefficient. 

Since the supersymmetric transformation relates fermions to bosons, there
is a nontrivial relationship between the coupling coefficients involving the 
scalars and the fermion. The supersymmetric transformation forces a 
relationship (namely the common
$k_{\mu \nu}$) on the Lorentz-violating terms of Eq.~(\ref{Lagrangian})
that is similar to the common mass and couplings that are a well-known 
consequence of supersymmetric theories. 

If one modifies the supersymmetric transformations of the Wess-Zumino 
model by adding new terms involving the coefficients $k_{\mu \nu}$ 
coefficients,
\begin{eqnarray}
\delta A&=&\overline{\epsilon}\psi\;, \nonumber \\
\delta B&=&i\overline{\epsilon}\gamma _5 \psi\;, \nonumber \\
\delta \psi&=&-(i\slasha{\partial}+ik_{\mu \nu}\gamma ^\mu\partial ^\nu)
(A+i\gamma _5B)\epsilon+(F+i\gamma _5G)\epsilon\;, \nonumber \\
\delta F &=& -\bar{\epsilon}(i\slasha{\partial}+ik_{\mu \nu}
\gamma^\mu \partial ^\nu)\psi\;, \nonumber \\
\delta G &=& \bar{\epsilon}(\gamma _5\slasha{\partial}
+k_{\mu \nu}\gamma_5\gamma^\mu \partial ^\nu)\psi \;,
\label{transformation}
\end{eqnarray}
one finds that the Lagrangian is invariant up to a total derivative.

The commutator of two supersymmetry transformations in 
Eq.~(\ref{transformation})
yields
\begin{eqnarray}
\left[ \de_1,\de_2 \right] 
= 2 i \ol\ep_1 \ga^\mu \ep_2 \prt_\mu 
+ 2 i k_{\mu\nu}\ol \ep_1 \ga^\mu \ep_2 \prt^\nu,
\end{eqnarray}
which involves the generator of translations.
A modified supersymmetry algebra therefore exists, and 
the lagrangian in Eq.~(\ref{Lagrangian})
provides an explicit example of an interacting model 
with both exact supersymmetry and Lorentz violation..

One can also show that a modification of the supersymmetry transformation 
in Eq.~(\ref{transformation}) cannot be modified, say by changing the 
transformation of the scalar fields $A$ and $B$, because any such modification
would not result in a closure of the supersymmetry algebra.
In fact, one can understand the modification of the Lagrangian and the 
supersymmetric transformations as a global substitution of the 
form $i\partial _\mu\to i\partial _\mu +ik_{\mu \nu}\partial ^\nu$.
The translation generator $P_\mu$ 
commutes with itself (first equation in Eq.~(\ref{poincare})) and satisfies
the first equation in Eq.~(\ref{susy}), so it then follows that
\begin{eqnarray}
&&\left [Q,P^2 \right ] =0\;.
\end{eqnarray}
Since the superpotential containing the mass and coupling terms
is unaffected by the Lorentz violation, 
analogues should exist for
various conventional results on supersymmetry breaking 
\cite{Wess:1974kz,O'Raifeartaigh:1975pr,Witten:1981nf}.

One consequence of the supersymmetric Lagrangian is the relationship between
the fermionic and scalar propagators. The fermionic propagator is
\begin{eqnarray}
{{i}\over {p_\mu (\gamma^\mu + k_{\mu \nu}\gamma ^\nu) - m}}\;.
\end{eqnarray}
Rationalizing this propagator by multiplying by 
$p_\mu (\gamma^\mu + k_{\mu \nu}\gamma ^\nu) + m$, one gets the denominator
\begin{eqnarray}
&&p^2+p^\mu p^\rho(k_{\mu \nu}\gamma^\nu \gamma _\rho + k_{\rho
\sigma}\gamma_\mu \gamma ^\sigma)\nonumber \\
&&\qquad +k_{\mu \nu}k_{\rho \sigma}p^\mu p^\rho
\gamma ^\nu \gamma ^\sigma -m^2 \;.
\end{eqnarray}
For $k_{\mu \nu}$ symmetric, the second term is $2k_{\mu \nu}p^\mu p^\nu$,
and the third term is 
$k_{\mu \nu}k^\mu _{\:\:\: \rho} p^\nu p^\rho$.
Consequently the scalar and fermion, which are related by supersymmetry, have
propagators with the same structure.

One might try to eliminate the Lorentz violation by a choice of coordinates
such that $x^\mu \to x^\mu +k^{\mu \nu} x_\nu$. However, the metric would then 
no longer have the usual Minkowski form, and the Lorentz violation is simply
moved into the new metric. The Lorentz violation of the theory is physical.

\section{CPT-Odd Lorentz Breaking}

It is well-known that a local Lorentz-invariant quantum 
field theory preserves the
combination $CPT$ where $C$ is charge conjugation, 
$P$ is parity, and $T$ is time 
reversal. If Lorentz-violating terms are 
added to the Lagrangian, however, 
then this result no longer needs to hold and nonlocality 
as well as CPT violation
is permitted. 

A $CPT$-violating term can be added to the Wess-Zumino 
model in the following way, 
\begin{eqnarray}
\cL &=&{1\over 2}\partial _\mu A\partial ^\mu A 
+{1\over 2}\partial _\mu B\partial ^\mu B 
+{1\over 2}i\overline{\psi}\slasha{\partial}\psi+{1\over 2}F^2+{1\over 2}G^2
\nonumber \\
&&+k_\mu(A\partial ^\mu B-B\partial ^\mu A)
-{1\over 2}k_\mu \overline{\psi}\gamma _5\gamma^\mu \psi 
\nonumber \\
&&+{1\over 2}k^2(A^2+B^2)\;.
\label{Lagrangian2}
\end{eqnarray}
The coefficient $k_\mu$ has mass dimension one, and a unique supersymmetry 
transformation 
can be established on 
dimensional grounds to be 
\begin{eqnarray}
\delta A&=&\overline{\epsilon}\psi\;, \nonumber \\
\delta B&=&i\overline{\epsilon}\gamma _5 \psi\;, \nonumber \\
\delta \psi&=&-(i\slasha{\partial}+\gamma _5\slasha{k})(A+i\gamma _5B)\epsilon
+(F+i\gamma _5G)\epsilon\;, \nonumber \\
\delta F &=& -\bar{\epsilon}(i\slasha{\partial}-\gamma _5\slasha{k})\psi\;,
\nonumber \\
\delta G &=& \bar{\epsilon}(\gamma _5\slasha{\partial}+i\slasha{k})\psi\;,
\label{transformation2}
\end{eqnarray}
and the Lagrangian again transforms into a total derivative.

The Lagrangian can be derived from the original Wess-Zumino model by 
applying a field redefinition,
\begin{eqnarray}
&&\psi \to e^{-i\gamma_5k\cdot x}\psi \nonumber \\
&&A\pm iB\to e^{\pm ik\cdot x}(A\pm iB)\;.
\label{redef}
\end{eqnarray}

The terms involving the coupling $k_\mu$ respect $C$ and $T$ but violate 
$P$, giving an overall $CPT$ violation.
Since parity is violated by the terms involving $k_\mu$ (both in the 
Lagrangian and in the transformations) one obtains a supersymmetric 
transformation that acts differently on the left- and right-handed states,
\begin{eqnarray}
\de \ps_L=(-i\slasha{\prt}+\slasha{k})(A+iB)\ep_R
+(F-iG)\ep_L, \nonumber \\
\de \ps_R=(-i\slasha{\prt}-\slasha{k})(A-iB)\ep_L
+(F+iG)\ep_R,
\end{eqnarray}
where the chiral components are defined as usual: $\psi _i=P_i\psi$,
$\epsilon _i=P_i\epsilon$, $P_{L/R}=(1\mp \gamma_5)/2$.

The mass and coupling terms in Eq.~(\ref{Lagrangian}) 
are not supersymmetric. The parity-odd mass and coupling
terms are also not consistent with supersymmetry. For example, the
parity-violating mass term
\begin{eqnarray}
m\left (i\overline{\psi}\gamma _5\psi+2AG-2BF\right )\;, 
\label{mass2}
\end{eqnarray}
cannot be added to the Wess-Zumino theory because it
is not invariant under the modified supersymmetry transformations in 
Eq.~(\ref{transformation2}). It is 
not surprising that they cannot be reconciled with supersymmetry 
since the mass and coupling terms do not respect the field redefinition.

In addition to the $k_{\mu \nu}$ and $k_\mu$-dependent terms described above,
the simple field content of the Wess-Zumino model admits additional
renormalizable Lorentz-violating term to its Lagrangian: 
$(A^2 \prt B \pm B^2 \prt A)$,
$\ph\ol\ps \ga_5\ga^\mu\ps$,
and $\ol{\ps}\si ^{\mu \nu}\prt ^\la \ps$. However, there do not appear to be 
supersymmetric interpretations for these terms.

\section{Superfield Formulation}

The elegant method of superspace~\cite{Salam:1974yz} allows one to combine all
the component 
fields of a supersymmetric multiplet into one superfield. By extending 
the four dimensions of spacetime to include fermionic dimensions as well, this
technique highlights the role of supersymmetry as a spacetime symmetry.
The fermionic
coordinates, $\theta$, that are added to the usual spacetime coordinates
highlight the role of supersymmetry as a spacetime symmetry. The
CPT transformation maps the components of a chiral superfield $\Phi$ into 
themselves, while the parity transformation alone maps left-chiral 
superfields into right-chiral superfields and vice versa.
Define 
\begin{eqnarray}
&&\ph = \frac 1 {\sqrt 2} (A + i B),
\quad
\cF = \frac 1 {\sqrt 2} (F - i G).
\end{eqnarray}
In terms of these complex scalars,
the left-chiral superfield appropriate for the model  
in Eq.~(\ref{Lagrangian}) is
\begin{eqnarray}
\Ph(x,\theta) &=& 
\ph (x) + \sqrt 2 \ol \theta \ps_L (x) 
+ \half \ol \theta (1 - \ga_5) \theta \cF(x)
\nonumber\\
&&
+\half i \ol \theta \ga_5 \ga^\mu \theta 
(\prt_\mu + k_{\mu\nu} \prt^\nu)\ph (x) 
\nonumber\\
&&
- \frac i {\sqrt 2} \ol \theta \theta \ol\theta 
(\slasha\prt + k_{\mu\nu}\ga^\mu\prt^\nu) \ps_L (x)
\nonumber\\
&&
- \frac 1 8 (\ol\theta\theta)^2 (\prt_\mu + k_{\mu\nu}\prt^\nu)^2 \ph (x).
\label{superfield}
\end{eqnarray}
Here, 
the subscript $L$ denotes projection 
with $\half (1 - \ga_5)$.
The lagrangian in Eq.~(\ref{Lagrangian})
can then be expressed as 
\begin{eqnarray}
\cL = \Ph^* \Ph\vert _D
+ \left( \half m \Ph^2\vert _F
+ \frac 13 g \Ph^3\vert_F + {\rm h.c.}\right),
\label{sfieldlag}
\end{eqnarray}
where the symbols $\vert_D$ and $\vert_F$
refer to projections onto the $D$- and $F$-type components 
of the (holomorphic) functions of $\Ph (x, \theta)$.
The theory can therefore be represented as an action in superspace.

A supersymmetry transformation on $\Ph (x, \theta)$
generated as $\de_Q \Ph (x,\theta) = - i \ol\ep Q \Ph(x,\theta)$ where
\begin{eqnarray}
Q&=& i \prt_{\ol \theta} - \ga^\mu \theta \prt_\mu
- k_{\mu\nu} \ga^\mu \theta \prt^\nu.
\label{q}
\end{eqnarray}
It is easy to check that the application of the supersymmetry
transformation to the resulting chiral superfield yields the transformation
given in Eq.~(\ref{transformation}), and that the supersymmetry algebra closes.
The superpotential is not modified and gives rise to the mass and coupling 
terms in Eq.~(\ref{Lagrangian}) 
in the usual way by projecting out the usual $F$-term component of the 
holomorphic functions of the chiral superfield $\Phi$.
The generators $Q$
and $P_\mu = i \prt_\mu$ satisfy
\begin{eqnarray}
&&\left [Q,P_\mu \right ] =0,
\quad
\left \{Q,\ol{Q}\right \}
=2\ga ^\mu P_\mu  + 2 k_{\mu\nu}\ga^\mu P^\nu,
\label{superalg}
\end{eqnarray}
which represents a new superalgebra explicitly containing the Lorentz violation
parameter $k_{\mu \nu}$.

\section{Conclusions}

Lorentz-violating extensions of 
the Wess-Zumino model have been found that remain exactly supersymmetric. 
Nontrivial relationships exist between terms in 
the Lagrangian involving scalars and terms involving fermions reminiscent of
the common masses and couplings in other supersymmetric theories.
The supersymmetry algebra closes in a modified way and the  
momentum generator continues to generate translations.
Therefore a representation of supersymmetry has been found that does not
respect the Lorentz invariance of the superPoincar\'e algebra.
Many of the appealing features of supersymmetry are preserved even if Lorentz
violation is allowed. 
One might extend the considerations discussed here to other representations of
supersymmetry. For example the vector superfield should have a straightforward
generalization to include Lorentz violation. 
One could then conceive of phenomenological models (like supersymmetric QED or
the Minimal Supersymmetric Standard Model) that include explicit terms
that break supersymmetry, Lorentz invariance, or both. 

The Lorentz symmetry has long been viewed by many as one which is 
unlikely to be broken. However, one 
should be wary of arguments that say since a dimensionless number 
is constrained to be very small, it must be exactly zero. 
The prominent recent example is the observed size of a ``cosmological 
constant'' compared to known particle physics scales. Furthermore 
large hierarchies exist between observed physical scales in particle physics: 
there is a large gap of at least six orders of magnitude between the electron 
mass and the heaviest neutrino, so we can confidently say that a ``desert'' 
unpopulated by massive states has already been confirmed experimentally.
Whether Lorentz violation exists and is characterized by very small 
dimensionless numbers or ratios is a question that can ultimately only be
decided by experiment.


\begin{theacknowledgments}
I would like to thank my collaborator V. Alan Kosteleck\'y.
This work was supported in part by the U.S.
Department of Energy under Grant No.~DE-FG02-91ER40661.
\end{theacknowledgments}


\bibliographystyle{aipproc}   

\bibliography{coralgables}

\hyphenation{Post-Script Sprin-ger}
\begin{thebibliography}{11}
\expandafter\ifx\csname natexlab\endcsname\relax\def\natexlab#1{#1}\fi
\providecommand{\enquote}[1]{``#1''}
\expandafter\ifx\csname url\endcsname\relax
  \def\url#1{\texttt{#1}}\fi
\expandafter\ifx\csname urlprefix\endcsname\relax\def\urlprefix{URL }\fi

\bibitem[Colladay and Kostelecky(1997)]{Colladay:1997iz}
Colladay, D., and Kostelecky, V.~A., \emph{Phys. Rev.}, \textbf{D55},
  6760--6774 (1997).

\bibitem[Colladay and Kostelecky(1998)]{Colladay:1998fq}
Colladay, D., and Kostelecky, V.~A., \emph{Phys. Rev.}, \textbf{D58}, 116002
  (1998).

\bibitem[Kostelecky and Samuel(1990)]{Kostelecky:1990nt}
Kostelecky, V.~A., and Samuel, S., \emph{Nucl. Phys.}, \textbf{B336}, 263
  (1990).

\bibitem[Kostelecky and Potting(1996)]{Kostelecky:1996qk}
Kostelecky, V.~A., and Potting, R., \emph{Phys. Lett.}, \textbf{B381}, 89--96
  (1996).

\bibitem[Berger and Kostelecky(2002)]{Berger:2001rm}
Berger, M.~S., and Kostelecky, V.~A., \emph{Phys. Rev.}, \textbf{D65}, 091701
  (2002).

\bibitem[Wess and Zumino(1974{\natexlab{a}})]{Wess:1974tw}
Wess, J., and Zumino, B., \emph{Nucl. Phys.}, \textbf{B70}, 39--50
  (1974{\natexlab{a}}).

\bibitem[Kostelecky and Lehnert(2001)]{Kostelecky:2001rh}
Kostelecky, V.~A., and Lehnert, R., \emph{Phys. Rev.}, \textbf{D63}, 065008
  (2001).

\bibitem[Wess and Zumino(1974{\natexlab{b}})]{Wess:1974kz}
Wess, J., and Zumino, B., \emph{Phys. Lett.}, \textbf{B49}, 52
  (1974{\natexlab{b}}).

\bibitem[O'Raifeartaigh(1975)]{O'Raifeartaigh:1975pr}
O'Raifeartaigh, L., \emph{Nucl. Phys.}, \textbf{B96}, 331 (1975).

\bibitem[Witten(1981)]{Witten:1981nf}
Witten, E., \emph{Nucl. Phys.}, \textbf{B188}, 513 (1981).

\bibitem[Salam and Strathdee(1974)]{Salam:1974yz}
Salam, A., and Strathdee, J., \emph{Nucl. Phys.}, \textbf{B76}, 477--482
  (1974).

\end{thebibliography}

\IfFileExists{\jobname.bbl}{}
 {\typeout{}
  \typeout{******************************************}
  \typeout{** Please run "bibtex \jobname" to optain}
  \typeout{** the bibliography and then re-run LaTeX}
  \typeout{** twice to fix the references!}
  \typeout{******************************************}
  \typeout{}
 }

\end{document}